\documentclass[journal]{IEEEtran}
\usepackage{cite}
\usepackage[pdftex]{graphicx}
\usepackage[cmex10]{amsmath}
\usepackage{array}
\usepackage{url}
\usepackage{textcomp}
\usepackage{hyperref}
\usepackage{color}
\usepackage{graphicx,subfigure}
\def\e{\begin{equation}}
\def\f{\end{equation}}
\def\l#1{\label{eq:#1}}
\def\r#1{(\ref{eq:#1})}
\begin{document}

\title{Decoupling of Two Closely Located Dipole Antennas by a Split-Loop Resonator}

\author{Masoud Sharifian Mazraeh Mollaei, Anna Hurshkainen, Sergey Kurdjumov, Stanislav Glybovski and Constantin Simovski
\thanks{M. S. M. Mollaei and C. Simovski are with Aalto University, FI-00076 Espoo, Finland}
\thanks{A. Hurshkainen, S. Kurdjumov and S. Glybovski are with ITMO University, 197101 St. Petersburg, Russia}
}



\maketitle

\begin{abstract}
In this letter, we theoretically and experimentally prove the possibility of the complete passive decoupling for two parallel resonant dipoles by a split-loop resonator.
Unlike previously achieved decoupling by a similar resonant dipole, this decoupling technique allows us to avoid the shrink of the operation band. Compare to previous work, simulation and measurement show 100\% enhancemnet of relative operation band from 0.2\% to 0.4\%.
\end{abstract}


\section{Introduction}

In many radio-frequency applications, antenna arrays consist of closely located dipoles and their decoupling is required. When the straightforward methods of decoupling
(screens or absorbing sheets) are not applicable, one often uses adaptive technique when the decoupling is achieved involving active circuitry -- operational amplifiers.
However, in multi-input multi-output (MIMO) systems and antenna arrays for magnetic resonance imaging (MRI) the passive decoupling is preferred \cite{1,2,3,4,5,6}.  The keenest situation
corresponds to compact arrays when the distance $d$ between two parallel dipole antennas is smaller than $\lambda/10$, where $\lambda$ is the wavelength in the operation band. Then, this gap is not sufficient in order to introduce an electromagnetic band-gap (EBG) structure or to engineer a defect ground state \cite{1,2,3}. For passive decoupling of the loop antennas
used in MRI radio-frequency coils one found specific technical solutions working for densely packed arrays (see e.g. in \cite{4}). As to dipole arrays,
the passive decoupling is realized either involving the strongly miniaturized (and challenging in its tuning) EBG structures \cite{5} or arrays of passive scatterers \cite{6}.
However, in both these cases the success was achieved when $d\approx \lambda/12$, whereas there is a strong need in dipole antenna arrays arranged with $d<\lambda/30$ \cite{1,7}.

A complete passive decoupling of two resonant dipole antennas 1 and 2 separated by an arbitrary gap $d$ (the minimal value of $d$ is restricted only by the requirement $d\gg r_0$, where
$r_0$ is the wire cross section radius) was suggested and studied theoretically and experimentally in \cite{8}. The decoupling is achieved by placing a similar resonant but passive dipole 3
(half-wave straight wire) in the middle of the gap. Then the electromotive force (EMF) induced by dipole 1 in dipole 2 is compensated by a part of the  electromotive force induced in dipole 2 by scatterer 3. Similarly, scatterer 3 when excited by dipole 2 compensates the EMF induced by dipole 2 in dipole 1.  This decoupling is complete, meaning that the
power flux from dipole 1 to dipole 2 (and vice versa) is substituted by the flux from these dipoles to the scatterer for whatever relations between currents in dipoles 1 and 2. They both can be active, one of them can be active whereas the other can be loaded at the center by a lumped load or can be shortcut, they remain decoupled. However, this decoupling is approximate in the meaning that the power flux between dipoles 1 and 2 is suppressed not completely. For practical applications it is enough to reduce the mutual power transmittance by 10-12 dB. However, the bandwidth of antennas may suffer of the presence of scatterer 3. This is the case of \cite{8}, when the band of the resonant lossless matching (when the antenna circuits are tuned at the decoupling frequency)
shrunk seven times due to passive dipole 3; and the bandwidth of the decoupling regime was as narrow as the resonance band.

The purpose of the present study is to find a decoupling scatterer for two dipole antennas separated by a gap $d<\lambda/30$ as compact and efficient as the
half-wave straight wire but more broadband. We will show that it can be achieved using an elongated split loop resonating at the same frequency as dipoles 1 and 2.
We called such the loop (depicted in Fig.~\ref{figA}) split-loop resonator (SLR).

\section{Theory of decoupling by a split-loop resonator}

\begin{figure}[htb!]
\centering
\includegraphics[width=0.75\linewidth]{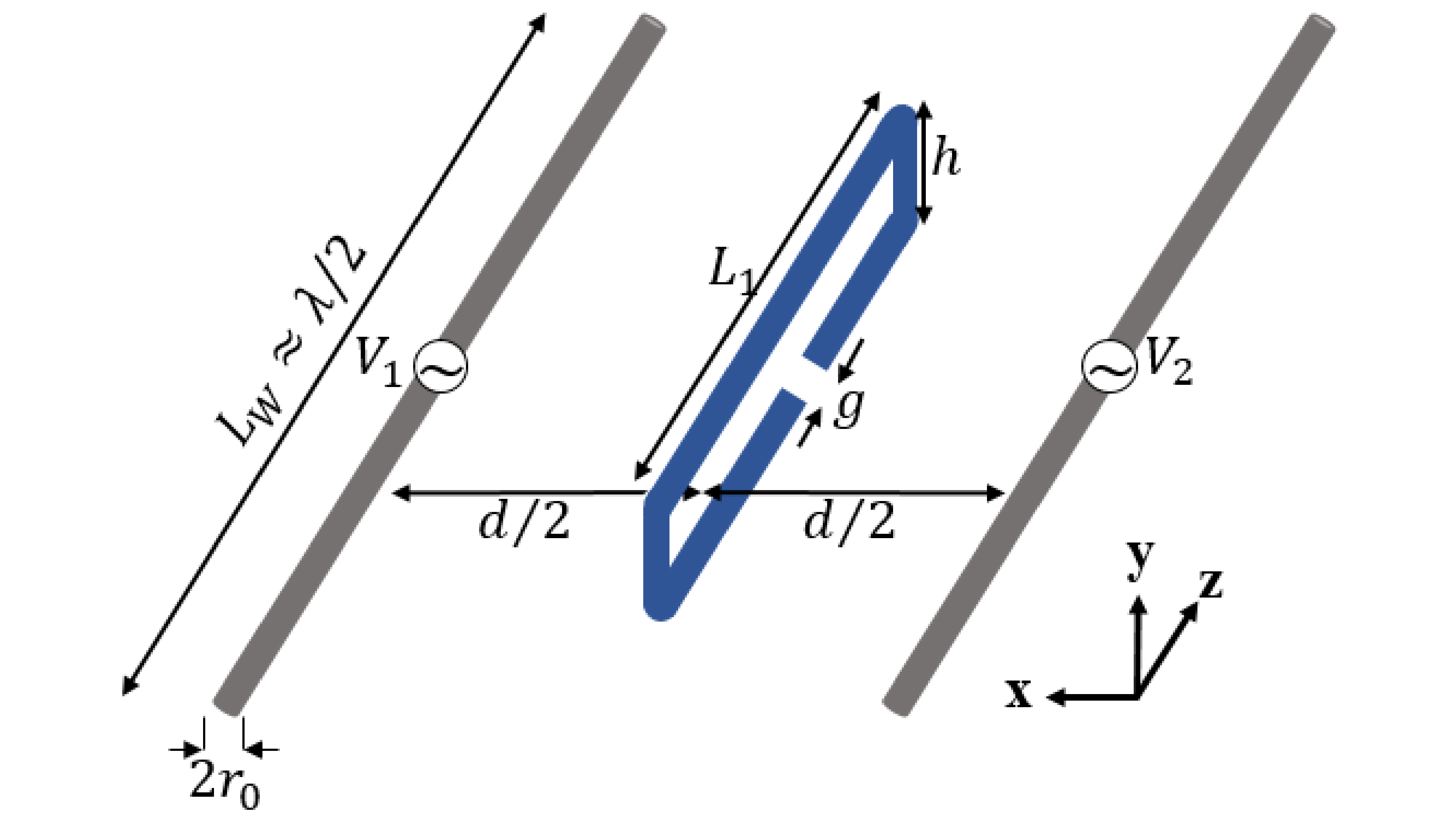}
\caption{A resonant SLR 3 located in the middle between antennas 1 and 2.}
\label{figA}
\end{figure}

\begin{figure}[htb!]
\centering
\includegraphics[width=0.85\linewidth]{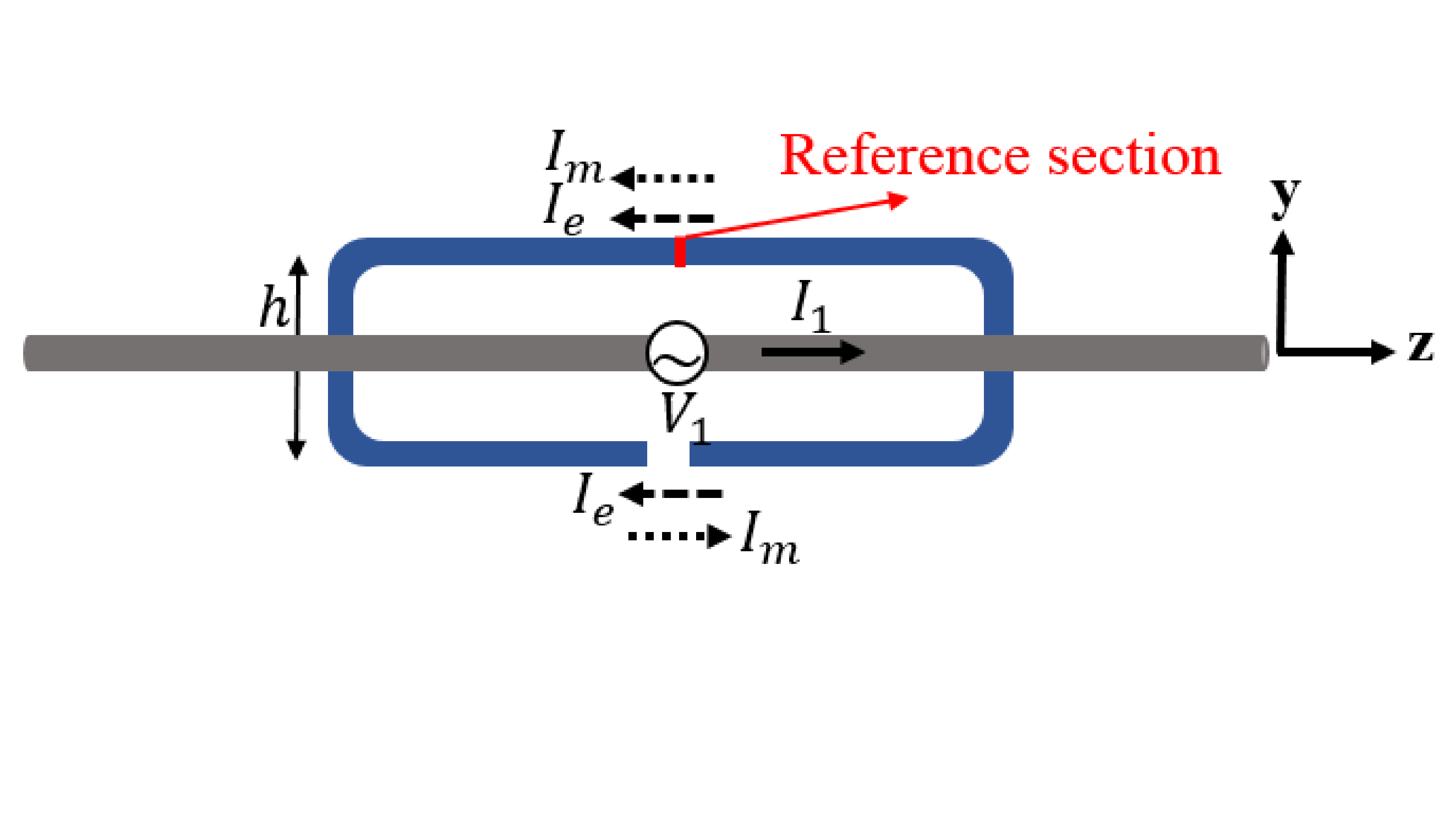}
\caption{The side view of the structure comprising an active dipole 1 driven by an external voltage $V_1$
and the passive SLR 3. Current $I_3$ induced in the SLR is the sum of the electric $I_e$ and magnetic $I_m$ modes.
At $(y=+h/2,\ z=0)$ $I_m=I_e=I_0/2$, at $(y=-h/2,\ z=0)$ $I_m=-I_e=-I_0/2$.}
\label{figB}
\end{figure}

Now let us prove that decoupling of dipoles 1 and 2 located in free space is  possible with an SLR symmetrically located between them.
Since the loop contour $C$ comprises the gap $g$ we may consider the SLR as a wire scatterer. The method of induced EMFs
is applicable to our SLR, as well as it was applicable to the dipole of our previous work \cite{8}. Therefore, the condition of
the complete decoupling expressed by formula (12) of work \cite{8}
\e
	{Z_{13}}^2 = {Z}{Z_M}  \l{cn12}\f
remains valid for the structure depicted in Fig.~\ref{figA}. Here $Z_M$ is mutual impedance between dipoles 1 and 2, $Z$ is the self-impedance of our SLR, and $Z_{13}=Z_{23}$
is its mutual impedance with antenna 1 or antenna 2. Both $Z$ and $Z_M$ are referred to the scatterer center -- reference section (RS)
located on the solid (top) side of the loop as shown in Fig. \ref{figB}.

In Fig.~\ref{figB} we depict the side view of our structure.
A primary source $V_1$ in the center of dipole 1 induces in our SLR 3 two current modes -- an electric one $I_e$
symmetric and an antisymmetric magnetic one $I_m$ with respect to the plane $y=0$. If the current in the reference section of the SLR i.e. at point
$(y=+h/2,\ z=0)$ is denoted as $I_0$ both modes have the same amplitude $I_0/2$ at this point, whereas they mutually cancel each other at the gap that can be approximated by point $(z=0, y=-h/2)$.
The distribution of the electric dipole mode along the SLR is similar to that in a straight wire and, therefore, can be approximated as (see e.g. in \cite{Balanis}):
\e
f_e(z)\equiv {I_e(z)\over {I_0/2}}={\sin k\left({L_l\over 2}-|z|\right)\over \sin {kL_l\over 2}}.\l{em11}\f
Contrary to the electric mode, the magnetic one is maximal at the vertical sides of the loop. This is so because these sides are shortcuts if our SLR is considered as
a two-wire line. This model of the loop results in the following approximation:
\e f_m(z)\equiv {I_m(z)\over {I_0/2}}= \pm {\cos k\left({L_l\over 2}-|z|\right)\over \cos {kL_l\over 2}}.
\l{em2}\f
Sign plus corresponds to the top side of the loop ($y=+h/2$), sign minus -- to the bottom side ($y=-h/2$).

Let us calculate the mutual impedance $Z_{13}$ between dipole 1 and SLR 3 applying the general formula of the induced EMF method:
\e
Z_{13}={1\over I_1}\int_C E_{13}(l) f_3(l)\, dl,
\l{zz13}\f
where $I_1$ is the current at the center of dipole 1, $E_{13}(l)$
is the tangential component of the electric field produced by this primary current at a point $l$ of the wire contour $C$ of scatterer 3, and $f_3(l)=f_e(l)+f_m(l)$ is the current distribution in
scatterer 3. Decomposition of the current induced in 3 onto the electric and magnetic modes allows us to
split the right-hand side of \r{zz13} into electric and magnetic mutual impedances formed by the coupling of the primary current $I_1$ with the electric and magnetic modes,
respectively. The antisymmetry of the magnetic mode results in two mutually cancelling EMFs induced in the top ($y=+h/2$) and bottom ($y=-h/2$) sides. In the vertical sides $E_{13}=0$.
Meanwhile, the equivalent EMFs corresponding to the electric mode sum up and \r{zz13} is simplified to
\e
Z_{13}={2\over I_1}\int\limits_{-L_l/2}^{L_l/2} E_{13}(z,y={h\over 2}) f_e(z)\, dz.
\l{zzz13}\f
This formula describes the mutual impedance of two effective dipoles one of which is dipole 1 length $L_w$, and the other one is
one half of the SLR, e. g. its top side $L_1$. 
The problem of $Z_{13}$ yields to the symmetric mutual coupling of two parallel dipoles of different lengthes.

Formulas for the mutual impedance of two parallel and symmetrically arranged dipoles are known. We will use the integral formula of
\cite{King1} which allows us to rewrite \r{zzz13} as
\e
Z_{13}={\eta\over 2\pi }\int\limits_{-L_l/2}^{L_l/2}f_e(z)
F(z,\delta)\, dz.
\l{Z1133}\f
Here $\eta=120\pi $ Ohm is free space impedance and it is denoted
$$
F(z,\delta)={e^{-jkr_+}\over r_+}+{e^{-jkr_-}\over r_-}-2\cos{kL_w\over 2} {e^{-jkr}\over r},
$$
$r=\sqrt{z^2+\delta^2}$, $\delta=\sqrt{(h/2)^2+(d/2)^2}$, and values $r_+$ and $r_-$ are distances from two ends of dipole 1 to the integration point:
$$
r_-=\sqrt{\left({L_w\over 2}-z\right)^2+\delta^2},\quad r_+=\sqrt{\left({L_w\over 2}+z\right)^2+\delta^2}.
$$
The result of the integration in \r{Z1133} can be presented in the closed form even in the present case $L_l\ne L_w$ (see e.g. in \cite{Elliot}). However, all known
representations of this result are too cumbersome. We will obtain a simpler expression for $Z_{13}$ suitable for our purpose.

Namely, let us assume that both dipole 1 and SLR 3 resonate at the same frequency and the decoupling holds in their resonance band.
The resonance of dipole 1 holds when $L_w\approx 0.496\lambda$ and in our SLR the loop inductance resonates with its capacitance.
Assuming the capacitance of the gap $g$ to be negligibly small (that is correct if $r_0\ll g\ll L_l$) we may calculate the inductance of our rectangular loop
using formulas of \cite{Kal}, and its capacitance -- using formulas of \cite{Yossel}. Choosing as an example
$L_w=500$ mm and $r_0=1$ mm (then the resonance band of dipoles 1 and 2 centered by the resonance frequency can be specified as 290-310 MHz)
we fit the resonance band of the SLR to that of the dipoles when $h=10$ mm and $L_l=290$ mm.

Since in this case $L_l$ is noticeably smaller than $\lambda/2$, the sinusoidal current distribution \r{em11} can be replaced
by its quadratic approximation $f_e(z)=1-\left({2z/L_l}\right)^2$. This formula seems to be rough, but it is even more accurate (at least when $L_l<\lambda/3$) than the commonly adopted sinusoidal approximation \r{em11} which is not smooth at $z=0$.
Substitution of the quadratic approximation into \r{Z1133} and variable exchanges $L_l/2\pm z\rightarrow \xi$ yield the right-hand side of this relation to a linear combination of
following integrals:
$$
J_1=\int\limits_{-L_l\over 2}^{L_l\over 2}{e^{-jk\sqrt{\xi^2+a^2}}\over \sqrt{\xi^2+a^2}}\, d\xi,
$$
$$
J_2=\int\limits_{-L_l\over 2}^{L_l\over 2}\xi{e^{-jk\sqrt{\xi^2+a^2}}\over \sqrt{\xi^2+a^2}}\, d\xi,
$$
and
$$
J_3=\int\limits_{-L_l\over 2}^{L_l\over 2}\xi^2{e^{-jk\sqrt{\xi^2+a^2}}\over \sqrt{\xi^2+a^2}}\, d\xi,
$$
where $a$ is a constant independ on $\xi$. Integrals of types $J_{1-3}$ were calculated using the simplest variant of the stationary phase formula (see e.g. in \cite{Ass}).
In all these integrals the stationary phase point $\xi$ centers the integration interval, whereas the contributions of the ends of this interval (points $\xi=\pm L_l/2$) cancel out in the final expression. This is not surprising because the dipole mode current nullifies at the edges of the SLR.

The stationary phase method is adequate because $L_w$ is large enough and function $F(z)$ is oscillating.
Skipping all involved but very simple algebra, the result takes form:
\e
Z_{13}\approx {\eta L_l\over 3\pi }\left[
{kL_we^{-jk\delta}\over 4\sqrt{2\pi} \delta}- \cos{kL_w\over 2}
{kL_we^{-jk\Delta}\over 2\sqrt{2\pi} \Delta}
\right].
\l{ZZZ}\f
Here it is denoted $\Delta=\sqrt{(L_w/2)^2+\delta^2}$. Further simplification results from the
resonant length of our dipoles $kL_w=\pi$. The term with $\Delta$ in \r{ZZZ} vanishes and we obtain:
\e
Z_{13}\approx {\eta L_le^{-jk\delta}\over 24\sqrt{2\pi}\delta}.
\l{ZZZZ}\f

Now, let us calculate the input impedance $Z$ of an individual SLR entering \r{cn12}. At frequencies near the resonance where the reactance is negligibly small,
the input impedance is equal (neglecting the Ohmic losses) to the radiation resistance $R_{SLR}$. This radiation resistance is a simple sum
of $R_{\rm el}$ -- that of a Hertzian dipole with effective length $L_{\rm eff}$ (see e.g. in \cite{Balanis})
\e
R_{\rm el}={\eta\over 6 \pi}(kL_{\rm eff})^2
\l{RE}\f
and $R_{\rm mag}$ -- that of a magnetic dipole with effective area $S_{\rm eff}$ (see e.g. in \cite{Balanis})
\e
R_{\rm mag}={8\pi\eta\over 3}(k^2S_{\rm eff})^2.
\l{RM}\f
Parameters $L_{\rm eff}$ characterizing the
distribution of the electric mode and $S_{\rm eff}$ (magnetic  mode) are easily found via simple integration of $f_e$ and $f_m$
that gives in our example case $L_{\rm eff}\approx L_l$ and $S_{\rm eff}\approx L_lh$.
%
Then, \r{RE} and \r{RM} for our example case give the radiation resistance of the resonant SLR $R_{SLR}=R_{\rm el}+R_{\rm mag}=\approx 70$ Ohm.
The resonant impedance of a half-wave dipole is also nearly equal $R_0=$70 Ohms \cite{Balanis}.
Therefore, it is reasonable to assume that the input impedance $Z$ of an individual SLR at frequencies near its resonance is
practically equal to that of the resonant dipole and can be approximated as $Z\approx R_0(1+\beta\gamma)$, where $\beta\approx 59$ and $\gamma=
(\omega-\omega_0)/\omega_0$ is relative detuning \cite{8}. Substituting this approximation for $Z$, \r{ZZZZ} for $Z_{13}$ and (14a) of \cite{8}
$Z_{M}\approx (\eta/24\pi kd)\exp{(-jkd)}$ into \r{cn12}, we obtain the decoupling condition as
\e
{R_0 \eta\over 24\pi kd}e^{-jkd}(1+\beta \gamma)=\left({\eta L_l\over 24}\right)^2{e^{-2jk\delta}\over 2\pi\delta^2}.
\l{Mac33}\f
In the case $h\ll d$ $\delta\approx d/2$ and complex exponentials cancel out that reduces \r{Mac33} to the simplest
equation from which we find the detuning $\gamma$ corresponding to the decoupling
\e
\beta \gamma=\left(\eta k L_l^2/d R_0\right)-1.
\l{Mac3}\f
For $d=3$ cm (in this case $h=d/3$) and $L_l=29$ cm \r{Mac3} yields $\gamma\approx 0.0423$ that implies the decoupling at the upper edge of the resonance band -- at 312.8 MHz.
Of course, this is an approximate decoupling, however, in our terminology it is complete since should be observed for whatever relations of currents in the active dipoles.


\section{Validation of the Theory and Discussion}

\begin{figure}[htb!]
\centering
\includegraphics[width=0.85\linewidth]{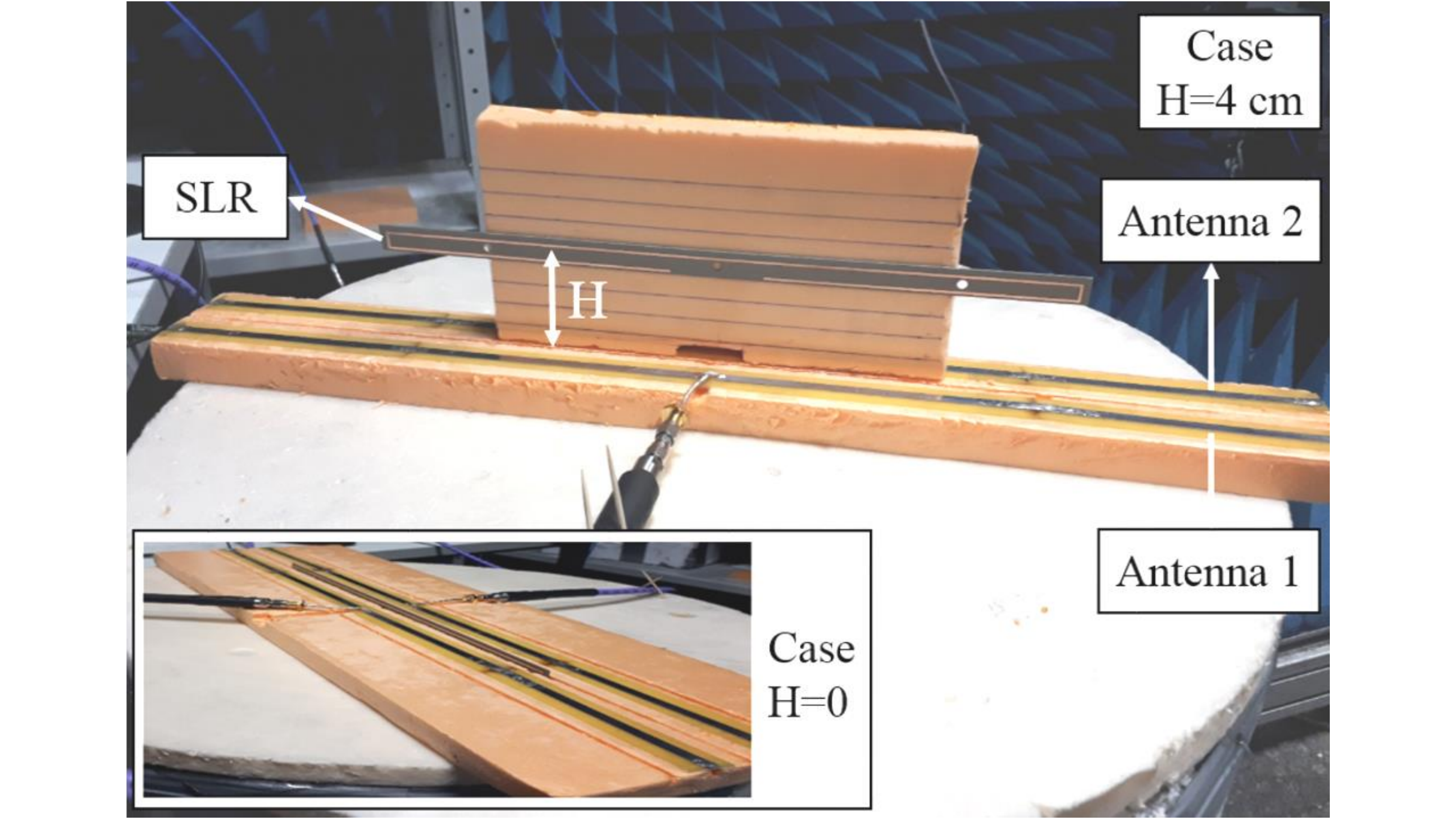}
\caption{Picture of the setup. The structure is supported by foam wrapped by paper. Slots in the vertical foam sheet show the height $H$ of the SLR
over the plane $y=0$. The complete decoupling corresponds to $H=0$.}
\label{figC}
\end{figure}

Numerical investigations of $S_{12}$ parameter calculated using CST Studio for dipoles 1 and 2 performed of a copper wire in absence and in presence of ideal matching circuits tuned at the
frequency of decoupling. Simulations were done in absence of our SLR (reference structure) and in its presence. The decoupling frequency was taken exactly equal to that
predicted by our theory (312.8 MHz) and the geometric parameters offering the decoupling at these frequency turned out to be surprisingly close to those predicted by our theory.
Namely, for $d=3$ cm the complete decoupling at frequency $312.8$ MHz is achieved with following design parameters of antennas and SLR: $r_0=1$ mm, $L_l=290.2$ mm, $L_w=500$ mm, $h=7$ mm.
Also, in these simulations we took $g=30$ mm that is not specified by the theory but satisfies its assumption $r_0\ll g\ll L_l$.
Simulations have confirmed our expectations about a broader band of both resonant matching and decoupling granted by an SLR compared to a resonant dipole \cite{8}.
The replacement of the decoupling dipole with an SLR enlarges the operation band from 0.2\% \cite{8} to 0.4\%.

For further validation we built an experimental setup pictured in Fig.~\ref{figC}. The setup is similar to that described in \cite{8}. The main difference is the
replacement of a straight wire by an SLR. Similar to \cite{8}, in this experiment we varied the height $H$ of our SLR over the plane of the dipoles $y=0$ ($H=0$ corresponds to the initial location of the SLR centered by this plane). Note that our model developed above does not prohibit the decoupling in the case when $H\ne 0$ and the magnetic mode is induced in the SLR. However, both CST simulations and this experiment have shown no complete decoupling for $H\ne 0$.

\begin{figure}[t!]
\centering
\subfigure[]{\includegraphics[width=9.7cm]{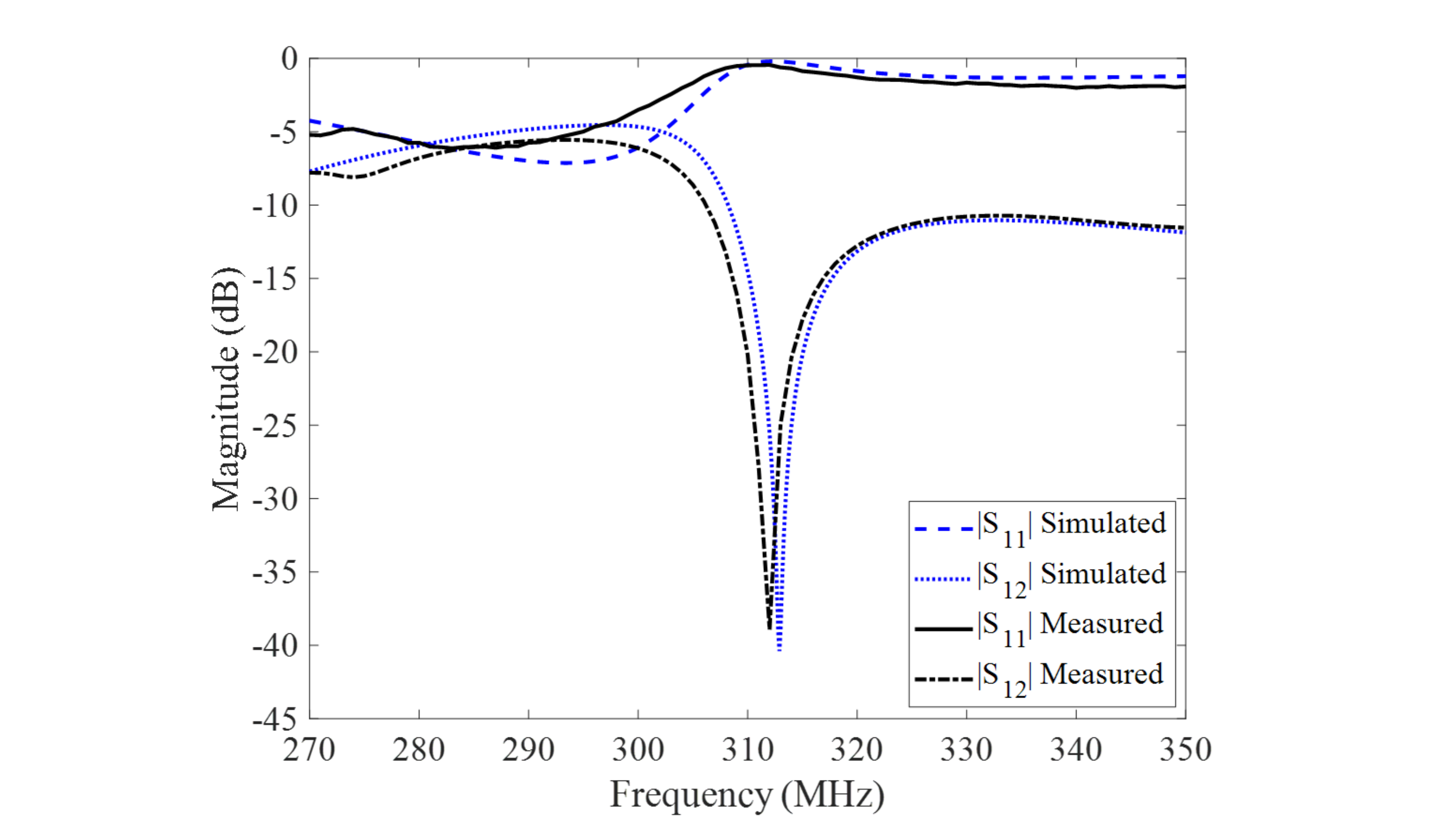}\label{Mas1}}
\subfigure[]{\includegraphics[width=9.7cm]{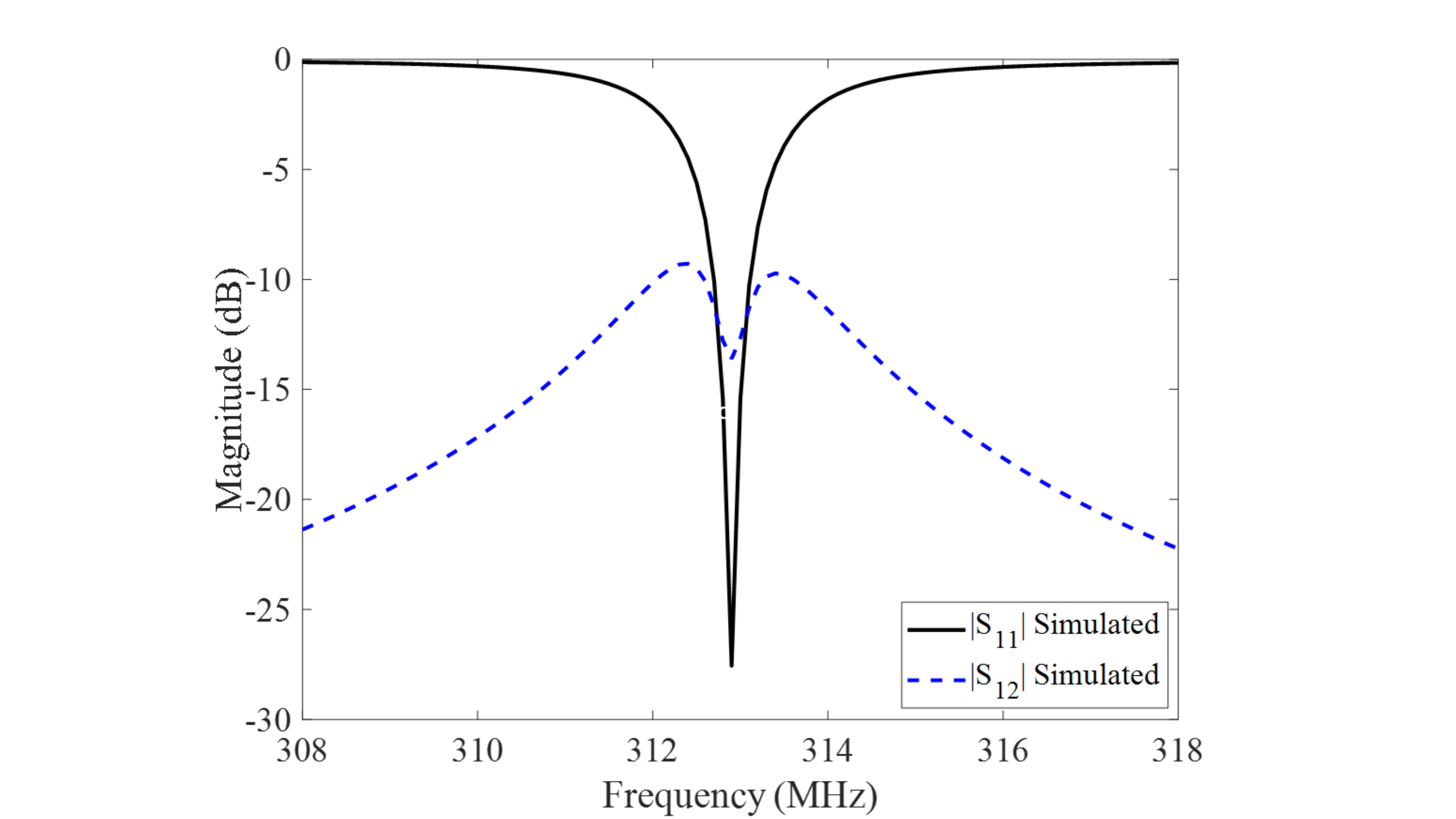}\label{Mas2}}
\caption{Frequency dependencies of $S_{11}$ and $S_{12}$ for the system of our
dipoles 1 and 2 decoupled by our SLR 3. (a) Mismatched case ,(b) matched case.}
\label{Masoud}
\end{figure}

Our experimental and numerical results are presented in Fig.~\ref{Masoud}. 
In both mismatched and matched regimes (Figs.~\ref{Mas1} and~\ref{Mas2}, respectively), minima of $S_{12}$ were simulated at $312.8$ MHz that is the indication of the complete decoupling.
Due to the difficulty of the tunable matching circuit, we measured the S-parameters only for the mismatched structure. 
Our measurements agree very well with simulations and can be considered as a confirmation of the theory. 

Our simulations for the matched case show that the insertion of SLR 3 decreases $S_{12}$ at 312.8 MHz by 10 dB (from -4 dB corresponding to the reference structure \cite{8} 
to -14 dB). This is probably sufficient for many applications. The operational band of the decoupled system can be defined as the minimal one of two bands -- that of the matching
(the band where $S_{11}\le -15$ dB using a lossless matching circuit) and that of the decoupling (the band where $S_{12}\le -10$ dB). In these definitions 
both bands of the matching and decoupling are equal to $1.3$ MHz. This band is much wider than that offered by a decoupling dipole in \cite{8} and this broadening is the main 
practical result. It follows from the fact that the extra mismatch due to the presence of the SLR at the distance $d/2$ from our antennas is not as high as the extra mismatch due to the 
presence of the dipole scatterer. We have not compared the simulation results of decoupling by passive SLR and dipole because the decoupling frequency is not the same for these cases; however, the enhancement of operating band is clear from comparison of $S_{11}$ in Fig.~\ref{Mas1} with $S_{11}$ in Fig. 3 of paper \cite{8}.   

\section{Conclusion}

In this Letter, we have theoretically and experimentally shown the complete (for whatever ratio of source voltages and currents) decoupling of two very closely located resonant dipoles is possible by adding a passive scatterer different from the similar dipole studied in our previous work. Decoupling can be granted by an elongated split loop, having the resonance in the same frequency band as the dipole antennas. The usefulness of this technical solution is the enlarged operation band. Also, this study opens the door to further search of decoupling scatterers.

\end{document}